\documentclass[letter]{aa} % for the letters 
\newcommand {\apgt} {\ {\raise-.5ex\hbox{$\buildrel>\over\sim$}}\ }
\newcommand {\aplt} {\ {\raise-.5ex\hbox{$\buildrel<\over\sim$}}\ }
\usepackage{graphicx}
\usepackage{txfonts}
\usepackage{longtable}
\begin{document}
   \title{Superluminal non-ballistic jet swing in the quasar \object{NRAO~150} revealed by mm-VLBI}

   \titlerunning{Superluminal non-ballistic jet swing in the quasar \object{NRAO~150} revealed by mm-VLBI}

   \author{I. Agudo  \inst{1,2} \and U. Bach \inst{2} \and T.~P. Krichbaum \inst{2} \and  A.~P. Marscher  \inst{3} \and I. Gonidakis \inst{4} \and P.~J. Diamond \inst{5}  \and M. Perucho \inst{2} \and W. Alef \inst{2} \and D.~A. Graham \inst{2} \and A. Witzel \inst{2}  \and J.~A. Zensus \inst{2} \and M. Bremer \inst{6} \and J.~A. Acosta-Pulido \inst{7} \and R. Barrena \inst{7}}

   \offprints{I. Agudo, \email{iagudo@iaa.es}}

   \institute{Instituto de Astrof\'{\i}sica de Andaluc\'{\i}a (CSIC),  Apartado 3004, E-18080 Granada, Spain
         \and
              Max-Planck-Institut f\"ur Radioastronomie, Auf dem H\"ugel, 69, D-53121, Bonn, Germany
         \and
              Institute for Astrophysical Research, Boston University, 725 Commonwealth Avenue, Boston, MA 02215, U.S.A.
         \and
              National and Kapodestrian University of Athens, Dept. of Astrophysics, Astronomy and Mechanics, GR-157 83 Athens, Greece
         \and
              University of Manchester, Jodrell Bank Observatory, Macclesfield, Cheshire SK11 9DL, U.K.
         \and
              Institut de RadioAstronomie Millim\'etrique, 300 Rue de la Piscine, 38406 St. Martin d'H\`eres, France
         \and
              Instituto de Astrof\'{\i}sica de Canarias, C/V\'{\i}a L\'actea s/n, E-38200, La Laguna, Tererife, Spain
           }

%   \date{Received September 15, 1996; accepted March 16, 1997}

% \abstract{}{}{}{}{}
% 5 {} token are mandatory

  \abstract
   {\object{NRAO~150}  --a compact and bright radio to mm source showing core/jet structure-- has been recently identified as a quasar at redshift $z=1.52$ through a near-IR spectral observation.}
   {To study the jet kinematics on the smallest accessible scales and to compute the first estimates of its basic physical properties,}
   {we have analysed the ultra-high-resolution images from a new monitoring program at 86~GHz and 43~GHz with the GMVA and the VLBA, respectively. An additional archival and calibration VLBA data set, covering from 1997 to 2007, has been used.}
   {Our data shows an extreme projected counter-clock-wise jet position angle swing at an angular rate of up to $\approx 11^{\circ}/\rm{yr}$ within the inner $\approx 31$~pc of the jet, which is associated with a non-ballistic superluminal motion of the jet within this region.}
   {The results suggest that the magnetic field could play an important role in the dynamics of the jet in \object{NRAO~150}, which is supported by the large values of the magnetic field strength obtained from our first estimates.
The extreme characteristics of the jet swing make \object{NRAO~150} a prime source to study the jet wobbling phenomenon.}

   \keywords{galaxies: active --
             galaxies: jets --
	     galaxies: quasars: general --
             galaxies: individual: \object{NRAO~150} --
	     radio continuum: galaxies --
             techniques: interferometric
               }

   \maketitle

%________________________________________________________________
\section{Introduction}
\label{int}

An increasing number of jets in active galactic nuclei (AGN) have been reported to show either regular or irregular swings of the innermost jet structural position angle in the plane of the sky (e.g., in \object{OJ~287}, Tateyama \& Kingham~\cite{Tat04}; in \object{3C~273}, Savolainen et al.~\cite{Sav06}; in \object{3C~345}, Lobanov \& Roland~\cite{Lob05}; in \object{BL~Lac}, Stirling et al.~\cite{Sti03}; in \object{S5~0716+71}, Bach et al.~\cite{Bac05}).
Time scales between 2 and 15 years and structural position-angle oscillations with amplitudes from $\sim 25^{\circ}$ to $\sim 45^{\circ}$ are typical for the reported cases.
We will call this phenomenon \emph{jet wobbling} hereafter.
Parsec scale AGN jet curvatures and helical-like structures also at larger distances from the central engine are also believed to be triggered by changes in direction at the jet ejection nozzle (e.g., in \object{3C~84}, Dhawan et al.~\cite{Dha98}).

The physical origin for the observed jet wobbling is still poorly understood. 
Among the various possibilities, regular precession of the accretion disk is frequently used for modeling at present.
Most AGN precession models are driven either by a companion super-massive black hole or another massive object inducing torques in the accretion disk of the primary (e.g., Lister et al.~\cite{Lis03}, for \object{4C~+12.50}; Stirling et al.~\cite{Sti03}, for \object{BL~Lac}; Caproni \& Abraham~\cite{CapAb04} for \object{3C~120}) or by the Bardeen-Peterson effect (e.g., Liu \& Melia~\cite{Liu02}; Caproni et al.~\cite{Cap04}).
However, other AGN scenarios that have yet to be explored extensively, such as the orbital motion of the jet nozzles (also involving binary systems)  or other kinds of more erratic disk/jet instabilities (e.g., similar to those thought to produce the quasi periodic oscillations [QPO] in X-ray binaries), can not be ruled out yet.
Note that, in support of these erratic instabilities, it is still under debate whether the observed jet wobbling is strictly periodic or not (see Mutel \& Denn~\cite{Mut05} for the case of \object{BL~Lac}).

There is still no paradigm to explain the phenomenon of jet wobbling in AGN, but it is rather likely that, as it is triggered in the innermost regions of the jets, it must be tied to fundamental properties of the inner regions of the accretion system.
Hence, there is ample motivation to study the jet wobbling phenomenon to place our understanding of the jet triggering region and the super--massive accretion systems on firmer ground.

VLBI observations at millimetre wavelengths are a powerful technique to image the innermost regions of AGN jets -which are self-absorbed at longer wavelengths- with the highest angular resolutions; $\sim 50\,\mu$as at 86~GHz (3.5~mm) and $\sim 0.15$\,mas at  43~GHz (7~mm).
Here, we report the discovery of an extreme case of jet swing in the quasar \object{NRAO~150}, through the first ultra-high-resolution VLBI set of images obtained from this source at 86~GHz and 43~GHz.
\object{NRAO~150} is a strong radio-mm source.
At radio wavelengths, on VLBI scales, \object{NRAO~150} displays a compact core plus a one-sided jet extending up to $r \apgt 80$~mas with a jet structural position angle (PA) of $\sim 30^\circ$ (e.g., Fey \& Charlot~\cite{Fey00}).
\object{NRAO~150} was not detected by the early optical surveys,  most probably due to obscuration through the
Milky Way (Galactic latitude $= -1.6^\circ$).
Almost 40~yr after its discovery at radio wavelengths (Pauliny-Toth, Wade \& Heeschen~\cite{Pau66}), \object{NRAO~150} has been identified as a quasar at redshift $z=1.52$ through a near-IR spectroscopic project (Acosta-Pulido et al. in prep.).

Throughout this paper we assume $H_{\circ} = 72$~km~s$^{-1}$~Mpc$^{-1}$, $\Omega_{m}=0.3$, and $\Omega_{\Lambda}=0.7$.
Under these assumptions, the luminosity distance of \object{NRAO~150} is $d _{L}=11025$~Mpc, 1~mas corresponds to $8.5$~pc in the frame of the source, and an angular proper motion of 1~mas/yr translates into a speed of $69.4~c$.
These parameters are used here for the first time to make quantitative estimates of the basic physical properties of the jet in \object{NRAO~150}.

%__________________________________________________________________
\section{Observations, images and their modelling}
\label{obs}

The data set presented here consists of 5 VLBI images at 86~GHz taken with the Global Millimetre VLBI Array (GMVA\footnote{http://www.mpifr-bonn.mpg.de/div/vlbi/globalmm}, and its predecessor, the CMVA) covering the time range from October 2001 to April 2004, and of 34 Very Long Baseline Array (VLBA) images taken at 43~GHz from June 1997 to February 2007. 
Among these 34 images, 20  of them (those obtained from June 1997 to May 2002) were taken as calibration measurements of the SiO maser monitoring program of \object{TX~Cam} (Diamond \& Kemball~\cite{Dia03}).
A new 8.4~GHz VLBA image obtained in June 2002 is also presented here.
Information about each single observation and its resulting image is provided in Table~\ref{obslog}.

The 86~GHz and the 43~GHz data were correlated at the Max-Planck-Institut f\"ur Radioastronomie and the VLBA correlators, respectively.
Subsequent phase and amplitude (including opacity) calibration were performed using standard procedures
within the AIPS software.
Further phase and amplitude self-calibration, as well as the final imaging, were performed through standard procedures within the Difmap software.

\begin{figure}
\centering
\includegraphics[width=8.5cm,clip]{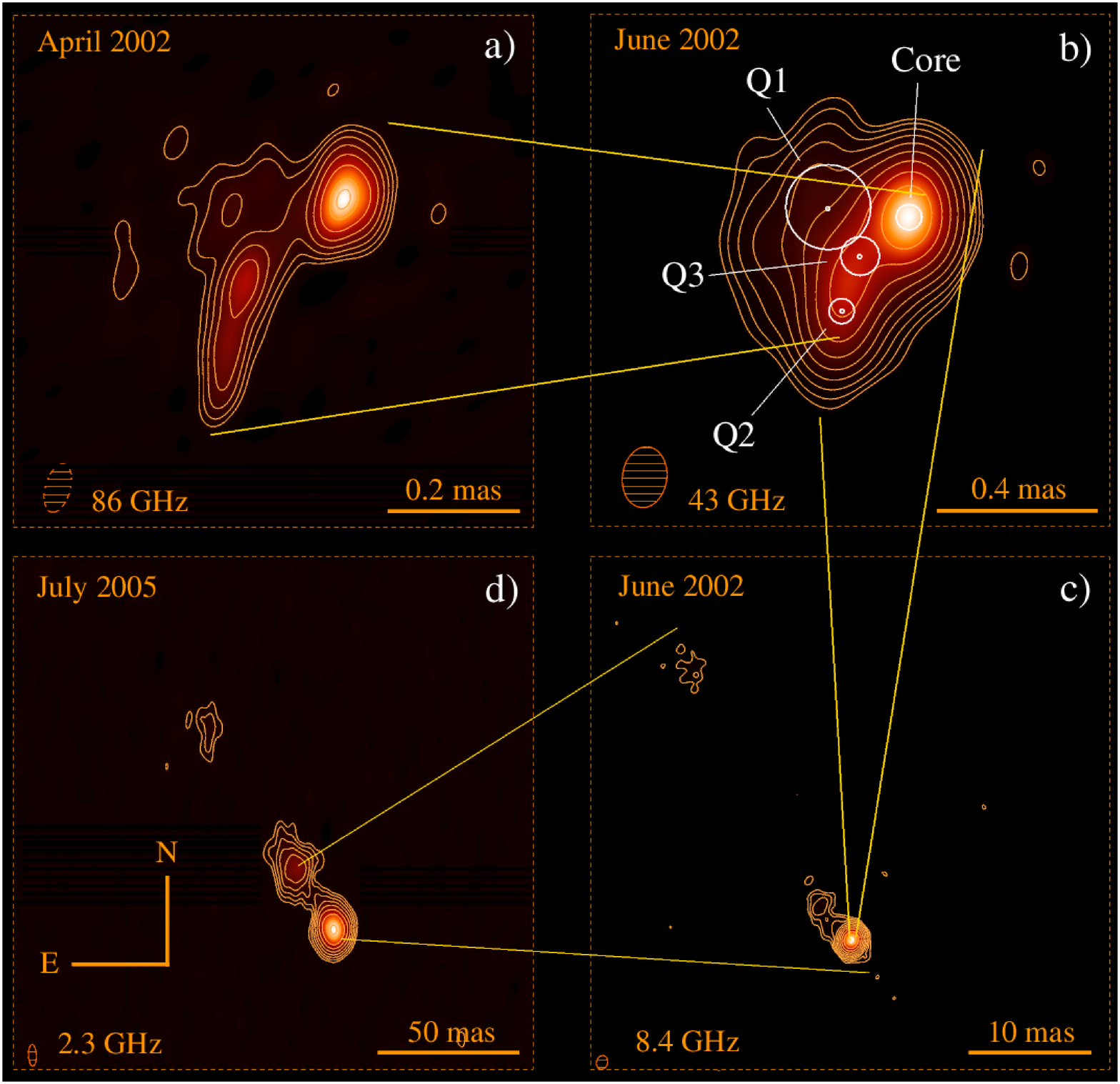}
\caption{VLBI images of \object{NRAO~150} at 86~GHz (a), 43~GHz (b), 8.4~GHz (c), and 2.3~GHz (d). 
The 2.3~GHz image was obtained by Kovalev et al.~(\cite{Kov07}).
In images a, b, c and d, the lower contours correspond to 1.0\%, 0.2\%, 0.15\% and 0.5\% of the corresponding image peak of 1.05~Jy/beam, 3.84~Jy/beam, 5.55~Jy/beam, and 0.69~Jy/beam, respectively.
The upper contours increase by factors of 2.
Other image parameters of the a), b) and c) maps are given in Table~\ref{obslog}.
The 2.3~GHz convolving beam is ($7.4\times2.6$)\,$\rm{mas}^{2}$ with major-axis position angle at $-2^{\circ}$.
The 86~GHz image illustrates the capability of the GMVA to achieve angular resolutions of up to 40~$\mu$as with dynamic ranges of up to 100:1 and notable image fidelity --as suggested by the matching 86~GHz and 43~GHz source structures--.
In the 43~GHz image, the circles are centred at the positions of the Gaussian model components. 
The radius of such circles symbolize the FWHM size of the corresponding model component.}
\label{misal}
\end{figure}

Examples of the resulting images at 86~GHz, 43~GHz, and 8.4~GHz --obtained in mid 2002-- are presented in Fig.~\ref{misal} together with a 2.3~GHz VLBA image obtained in July 2005 by Kovalev et al. (\cite{Kov07}).
At the four observing frequencies, the images show the typical core plus one sided jet structures of radio-loud quasars.
However, the orientation of the sub-mas scale jet (Fig.~\ref{misal}-a and b) differs by $>100^{\circ}$ with respect to that at larger scales (Fig.~\ref{misal}-c and d).
This suggests a bent structure of the inner jet oriented within a very small angle to the line of sight.

The series of 43~GHz images in Figure~\ref{im43GHz} shows evidence of motion in the North-West to South-East direction.
More intriguing is the fact that this motion is accompanied by an evident counter-clock-wise rotation of the jet on the plane of the sky, which will be the subject of discussion hereafter.

Difmap was also used to model the images of \object{NRAO~150} as sets of 3 or 4 circular-Gaussian intensity distributions (model components, e.g., Fig.~\ref{misal}).
Elliptical Gaussian fits were also tested, but they did not improve the results.
Only the minimum number of model components describing jet regions  considerably brighter than the image-background rms and at close positions during contiguous epochs were allowed.
All fitted model components resulted to have flux densities $S_{\rm{comp}}(86\,\rm{GHz})>90$\,mJy,  $S_{\rm{comp}}(43\,\rm{GHz})>50$\,mJy,  $S_{\rm{comp}}(8.4\,\rm{GHz})>60$\,mJy.
This characterized reliably each of the most significant jet regions by its flux density, relative position, and a FWHM size measurement at each observing epoch.
Table~\ref{mfits} lists the results from such fits, using the position of the (typically) brightest modelled Gaussian component --the VLBI core-- as reference.

%______________________________________________________________
\section{Superluminal motion and non-ballistic jet swing}
\label{kin}

\begin{figure*}
\centering
\includegraphics[width=17.5cm,clip]{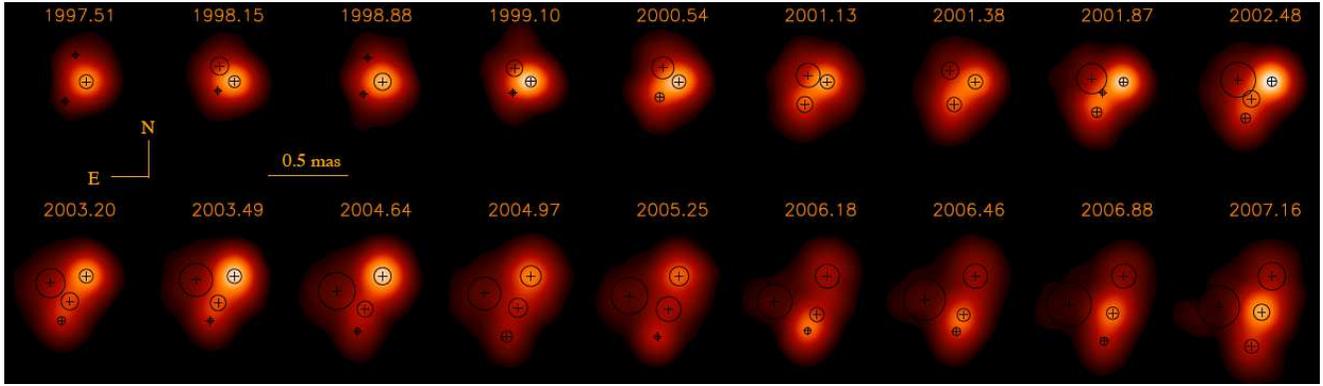}
\caption{Selection of 18 of the 34 \object{NRAO~150} images obtained from 1997 to 2007 with the VLBA  at 43~GHz. 
All images have been convolved with the same circular beam of 0.16~mas. 
The common intensity scale ranges from 0.025~Jy/beam to 4.128~Jy/beam.
The positions of the fitted Gaussians are indicated by the black crosses, whereas the black circles (of radius equal to the $FWHM$ of each Gaussian) symbolize their size.
A movie corresponding to these images can be downloaded from $<$http://www.iaa.es/$\sim$iagudo/research/NRAO150/NRAO150VLBA43GHznc.avi$>$.}
\label{im43GHz}
\end{figure*}

\begin{figure*}
\centering
\includegraphics[width=17.5cm,clip]{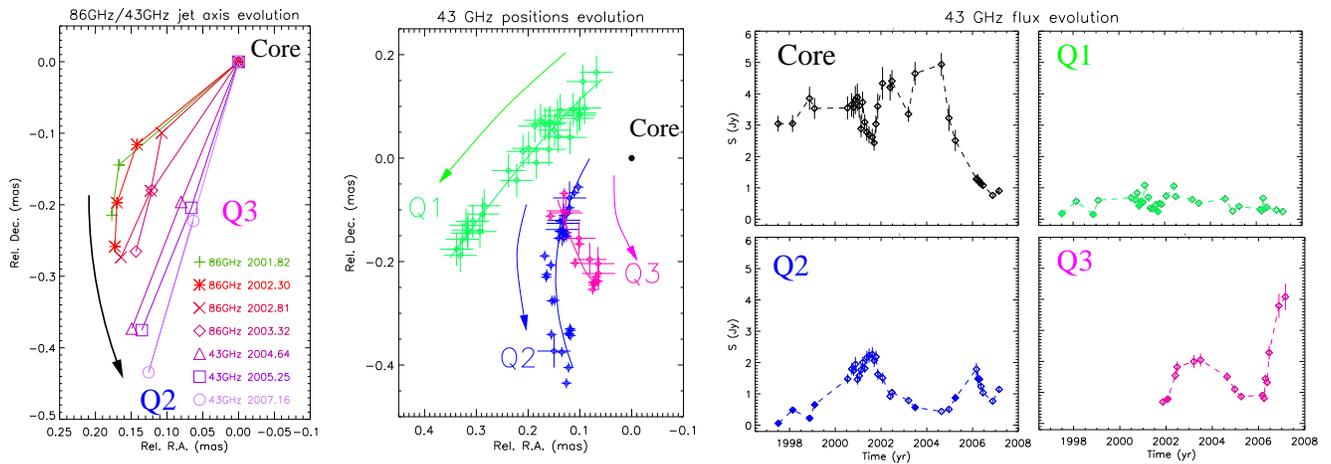}
\caption{From right to left, 86~GHz and 43~GHz inner jet axis evolution, projected trajectories of the 43~GHz model components, and their corresponding 43~GHz flux evolution. The crosses in the middle plot symbolize the position errors (computed using the prescriptions by Jorstad et al.~\cite{Jor05}), whereas the superimposed bent lines represent the fitted trajectories of each model component (labelled in the Figure). Filled diamonds in the flux evolution plots symbolize upper limits (see Table~\ref{mfits}).}
\label{fitsplot}
\end{figure*}

The 43~GHz jet structure evolution is illustrated in Fig.~\ref{im43GHz}, whereas Fig.~\ref{fitsplot} shows the 86~GHz and 43~GHz inner jet axis evolution, the projected trajectories of the three identified model components Q1, Q2, and Q3 with respect to the core position, and the total flux evolution of these components.
We modelled these trajectories, taking into account their curvatures, by fitting second order polynomials to them (e.g., as in Homan et al.~\cite{Hom01} and Jorstad et al.~\cite{Jor05}); see the results in Table~\ref{kintab}.
The mean measured proper motions are $(0.047\pm0.002)$~mas/yr, $(0.042\pm0.001)$~mas/yr, and $(0.033\pm0.002)$~mas/yr, for Q1, Q2 and Q3, with corresponding superluminal apparent speeds ($\beta_{\rm{app}}^{\rm{obs}}$) of $(3.26\pm0.14)~c$, $(2.85\pm0.07)~c$, and $(2.29\pm0.14)~c$, respectively.
Fig.~\ref{fitsplot} shows that the motion of features Q1, Q2 and Q3 outwards from the core is accompanied by counter-clock-wise rotation of their position vector, projected onto the plane of the sky, with extremely fast angular speeds $<\dot{\Theta}(\rm{Q1})> =(10.7\pm0.7)^{\circ}/\rm{yr}$, $<\dot{\Theta}(\rm{Q2})> =(5.8\pm0.3)^{\circ}/\rm{yr}$ and $<\dot{\Theta}(\rm{Q3})> =(8.0\pm0.5)^{\circ}/\rm{yr}$.

These are the first superluminal proper motions reported within the innermost 0.5~mas of the jet in \object{NRAO~150}.
We should also emphasize that the three fitted trajectories are bent and non radial, which implies that the motion of all the jet features are non-ballistic up to a projected distance from the core of $\sim0.5$~mas.
Assuming a jet viewing angle $\phi \approx 8^{\circ}$ (see Section~\ref{phys-par}), the jet features are implied to behave non-ballistically up to a deprojected distance from the core $\approx 31$~pc.

Similar results, in terms of the magnitude of projected angular speeds and non-ballistic nature, are obtained if either Q1 or Q2 is selected as the kinematic centre of the source.

Note also that when the mean projected velocity vectors of Q1, Q2 and Q3 are decomposed in their radial (with respect to the core) and non-radial (i.e. normal, $\beta^{\rm{obs}}_{\rm{app\_norm}}$) components, the later are $(2.71\pm0.21)$~$c$, $(1.39\pm0.21)$~$c$, and $(2.01\pm0.14)$~$c$, respectively.
Thus, we have shown here an extreme case of \emph{superluminal non-ballistic swing} in the jet of an AGN.
This is unusual because superluminal motions are reported in most cases in the radial direction relative to the core only.
Recently however, a growing number of AGN jets exhibiting bent model-component trajectories have been reported (e.g., Homan et al.~\cite{Hom01}; Jorstad et al.~\cite{Jor05}).
\object{NRAO~150} seems to be an extreme case, which reflects the remarkable non-ballistic nature of its jet. 
Both this non-ballistic nature and projection effects can explain why the trajectories of Q1 and Q3 do not cross with the one of Q2 when they are extrapolated back (see Fig.~\ref{fitsplot}). 

%______________________________________________________________
\section{Jet physical parameter estimates}
\label{phys-par}

Figure~\ref{fitsplot} shows the total flux density evolution of the model components.
Dramatic increases are displayed by Q2, at $\sim 2000$ and $\sim 2005$, and by Q3, at $\sim 2002$ and $\sim 2006$. 
Interestingly, a pronounced dip in the core light curve happens shortly before each of the dramatic Q2 flux peaks. 
These changes can be explained by changing Doppler boosting factors due to either a jet feature propagating through a bent underlying jet, or to the propagation of such a bend itself along the jet.

The ``synchrotron-self-Compton" (SSC) Doppler factor of the source ($\delta_{\rm{SSC}}$) can be constrained by comparing the observed  and the expected first order self-Compton X-ray flux density with the flux density of the source ($S_{\nu_{m}}$) at the synchrotron turnover frequency ($\nu_{m}$; e.g., Marscher~\cite{Mar83}; Agudo et al.~\cite{Agu06}).
Between August 1990 and February 1991,  \object{NRAO~150} displayed flux densities of $S_{1~keV}=(0.35\pm0.09)$~$\mu$Jy at 1~keV. 
ROSAT measured a flux $F_{0.1-2.4~keV}=(0.22\pm0.06)\times10^{-11}$~erg~cm$^{-2}$~s$^{-1}$ corrected for Galactic absorption (assumed a photon index $\Gamma=-1.7\pm0.1$) from \object{1RXS~J035930.6+505730} (Voges et al.~\cite{Vog00}), which position is fully consistent with the one of \object{NRAO~150} even at the level of the ROSAT 1-$\sigma$ position error. 
No other X-ray source or bright radio-mm source is found within 1~arcminute of the position of \object{NRAO~150}.
At $\nu_{m}\approx 37$~GHz, the flux density  $S_{\nu_{m}}=(2.59\pm0.21)$~Jy (see Reuter et al.~\cite{Reu97}; Voges et al.~\cite{Vog00}; Chen et al.~\cite{Che01}; Ter\"asranta et al.~\cite{Ter05}).
Given that the period between 1990 and 1997 was characterized by a low activity phase and that the VLBI structure of the source was dominated by the core in the 1990s, we assume that the bulk of the 37~GHz flux measured at the beginning of the 1990s was radiated from the core.
By taking into account these measurements, the average 43~GHz size of the core in our VLBI maps from 1997 to 2007, $<\theta>=(0.10\pm0.03)$~mas, and the typical optically thin synchrotron spectral index $\alpha \approx -0.5$ at the beginning of the 1990s, as well as the assumptions and approximations outlined in Agudo et al.~(\cite{Agu06}), we obtain $\delta_{\rm{SSC}}\approx 6$.

By using expression (2) in Marscher~(\cite{Mar83}) and the outlined measurements for the core ($S_{\nu_{m}} \approx 2.59$~Jy, $\nu_{m}\approx 37$~GHz, $\theta \approx 0.10$~mas, $\delta_{\rm{SSC}} \approx 6$ and $\alpha \approx -0.5$), the magnetic field intensity in the jet can be estimated.
In this way we find $\rm{B} \approx 0.7$~G, which is rather large for an AGN jet with the X-ray emission produced by SSC (e.g., Marscher et al.~\cite{Mar79}; Ghisellini et al.~\cite{Ghi98}).

Using now both $\beta_{\rm{app}}^{\rm{obs}}(Q1)$ as an estimate of the jet plasma speed and $\delta \approx \delta_{\rm{IC}}$ in the definitions of $\delta$\footnote{$\delta=[\gamma(1-\beta cos \phi)]^{-1}$ where $\gamma=(1-\beta^2)^{-1/2}$ is the Lorentz factor, $\beta$ is the speed in units of the speed of light and $\phi$ is the angle between the direction of the flow and the line of sight} and the apparent proper motion of the fluid, $\beta_{\rm{app}}$\footnote{$\beta_{\rm{app}}=(\beta sin \phi)/(1-\beta cos \phi)$}, we obtain $\gamma \approx 4$ and $\phi \approx 8^{\circ}$.

Note that all these estimates should still be taken as first approximations, since the used measurements are not as accurate and simultaneous as needed.
The results from a coordinated multi-waveband campaign organized in February 2007 will allow better and updated estimates.
In addition, the nature of Q1, Q2 and Q3 is still not known.  
They might be related to pattern speeds of propagating Kelvin-Helmholtz perturbations, which must have smaller speeds than those of the fluid in which they propagate (e.g., Perucho et al.~\cite{Per06}).
In contrast, they could be shock waves propagating in a bent underlying jet. 
In this case, the speed of the shocks must be an upper limit of the plasma speed. 
Nevertheless, our estimates of $\gamma$ and $\phi$ are consistent with those expected from the innermost regions of a quasar like \object{NRAO~150}, which extreme (projected) jet misalignment implies a very small viewing angle.

%________________________________________________________________
\section{Summary and conclusions}
\label{concl}

We have reported the results from the first 86~GHz and 43~GHz VLBI monitoring program of the recently identified quasar \object{NRAO~150}.
Our observations reveal
{\it a)} a large projected misalignment of the jet by $>100^{\circ}$ within the inner 0.5~mas to 1~mas from the core, 
{\it b)} an extremely fast counter-clockwise rotation of the projected jet axis at a rate of $\sim6^{\circ}/\rm{yr}$ to $\sim11^{\circ}/\rm{yr}$, 
{\it c)} non-ballistic superluminal motions with mean speeds from 2.3~$c$ to 3.3~$c$ within the inner 0.5~mas from the core (deprojected distance $\approx 31$~pc), 
{\it d)} transverse (non-radial) speeds of 2.7~$c$, 1.4~$c$, and 2.0~$c$ for Q1, Q2 and Q3, respectively, 
{\it e)} an extreme case of \emph{superluminal non-ballistic jet swing}, and
{\it f)} the first approximations to quantitative estimates of the basic physical properties of the jet in \object{NRAO~150}: $\delta \approx 6$, $B\approx 0.7$\,G, $\gamma \approx 4$, and $\phi \approx 8^{\circ}$.

Whereas the ultimate origin of the jet swing must be an intrinsic change of the direction of the inner jet axis (either caused by changes produced at the injection region or by interaction with the medium surrounding the jet), possible causes for the non-ballistic nature of the emitting flow are either a small inertia of the jet compared to that of the impacted ambient medium, or by the fact that we are observing a jet instability propagating downstream. 
However, the superluminal non-ballistic motion of the jet features in \object{NRAO~150} is perhaps too fast and systematic along tens of parsecs (deprojected, see above) to be induced solely by the ambient medium. 
In addition, the perturbation that must be produced either by the impact with such medium or by the changes at the injection region should imply the growth of disruptive instabilities (e.g., Perucho et al.~\cite{Per06}), in contrast to the remarkable collimation of the jet up to the kiloparsec scale (see Fig.~1). 
Mizuno et al.~(\cite{Miz07}) have shown that a magnetic field with the appropriate configuration helps keeping a jet collimated against the growth of instabilities. 
Hence, this possibility, together with the high value for the magnetic field intensity estimated in Section~\ref{phys-par}, suggests that the magnetic field could play an important role in the dynamics of the jet in \object{NRAO~150}. 

It is still unclear whether the reported change of the direction of ejection in \object{NRAO~150} is related to a regular (strictly periodic or not) behaviour or to a single event.
This, together with the still unknown nature of the moving jet features (either propagating curvatures or shocks in a bent jet), does not allow us to specify the ultimate origin of this phenomenon in the source.
This will be the matter of future studies.
Nevertheless, the extreme characteristics of the jet swing make \object{NRAO~150} a prime source for future studies of the origin of jet wobbling.

%______________________________________________________________
\begin{acknowledgements}

I. A. has been supported in part by an I3P contract with the Spanish ``Consejo Superior de Investigaciones Cient\'{i}ficas" and in part by a contract with the German ``Max Planck institute f\"ur Radioastronomie" (through the ENIGMA network, contract HPRN-CT-2002-00321), which were funded by the EU.
The GMVA is operated by the MPIfR, IRAM, OSO, MRO, and NRAO.
We thank the staff of the participating observatories for their efficient and continuous support.  
We acknowledge Y. Y. Kovalev, who provided the data for Fig.~\ref{misal}-d. 
This paper is partially based on observations carried out with the MPIfR 100~m Effelsberg Radio Telescope, the IRAM Plateau de Bure Millimetre Interferometer, the IRAM 30~m Millimetre Telescope, the Onsala 20~m Radio Telescope, the Mets\"ahovi  14~m Radio Telescope, and the VLBA. 
IRAM is supported by INSU/CNRS (France), MPG (Germany) and IGN (Spain).
The VLBA is an instrument of the NRAO, a facility of the National Science Foundation of the U.S.A. operated under cooperative agreement by Associated Universities, Inc. (U.S.A.).
This research has made use of the United States Naval Observatory (USNO) Radio Reference Frame Image Database (RRFID).
\end{acknowledgements}

%______________________________________________________________

\Online

\begin{table*}
\centering
\caption[]{86~GHz and the 43~GHz image information.
Given are the observing epochs, the total
integration times ($t_{\rm{int}}$), the observing frequency
bandwidth ($\Delta \nu_{\rm{obs}}$), the number of bits used for the
signal digitalization sampling, the FWHM minor and mayor
axes of the restoring beam and its orientation angle, the
integrated total flux densities, the peak flux density, and the noise levels
of the resulting images.}
\begin{flushleft}
\begin{tabular} {lccccccccc}
\hline\noalign{\smallskip}
Epoch & $t_{\rm{int}}$ &
$\Delta \nu_{\rm{obs}}$ & Bits & \multicolumn{3}{c}{Beam} & $S_{\rm{int}}$ & $S_{\rm{peak}}$ & Noise\\
 $[$y] &   [min]         &    [MHz]    &     &
$\mu$as  &  $\mu$as  &   $^{\circ}$  &      [Jy]    &    [Jy/beam]   &    [mJy/beam]      \\
\hline\noalign{\smallskip}
\hline\noalign{\smallskip}
\multicolumn{10}{c}{86~GHz images}\\
\hline\noalign{\smallskip}
2001.82  & 169  & 128  & 1  &  40 &  107 &  -21.1 & 3.38 & 1.22 & 3.3 \\
2002.30  & 293  & 128  & 1  &  38 &   77 &  -18.7  & 2.85 & 1.08 & 1.0 \\
2002.81  & 228  & 128  & 2  &  41 &   80 &  -15.6  & 3.82 & 1.30 & 2.4 \\
2003.32  &  98  & 128  & 2  &  75 &  221 &  -30.1  & 2.13 & 1.09 & 5.7 \\
2004.29  & 241  & 128  & 2  &  40 &   82 &  -14.1  & 4.57 & 1.03 & 1.6 \\
\hline\noalign{\smallskip}
\multicolumn{10}{c}{43~GHz images}\\
\hline\noalign{\smallskip}
1997.51  &  60  &  32  & 1  &  123 &  245 & -34.6 & 3.27 & 2.43 & 4.8  \\
1998.14  &  60  &  32  & 1  &  112 &  251 &  -9.4 & 4.06 & 2.68 & 4.8  \\
1998.88  &  60  &  32  & 1  &  154 &  230 &   1.1 & 4.25 & 3.04 & 3.6  \\
1999.10  &  60  &  32  & 1  &  187 &  292 &   1.1 & 4.78 & 3.80 & 3.8  \\
2000.54  &  20  &  16  & 2  &  129 &  241 & -19.0 & 5.73 & 3.30 & 4.0  \\
2000.72  &  20  &  16  & 2  &  132 &  244 & -19.0 & 6.04 & 3.39 & 3.5  \\
2000.80  &  20  &  16  & 2  &  132 &  266 & -21.3 & 6.18 & 3.42 & 3.4  \\
2000.88  &  20  &  16  & 2  &  242 &  356 & -36.7 & 6.18 & 4.75 & 4.4  \\
2000.97  &  20  &  16  & 2  &  205 &  329 &  23.4 & 5.92 & 3.85 & 4.5  \\
2001.05  &  20  &  16  & 2  &  208 &  340 &  13.8 & 5.72 & 3.85 & 4.1  \\
2001.13  &  20  &  16  & 2  &  206 &  268 &   5.7 & 5.74 & 3.40 & 3.7  \\
2001.20  &  20  &  16  & 2  &  195 &  261 &   8.8 & 5.73 & 3.35 & 5.8  \\
2001.29  &  20  &  16  & 2  &  225 &  275 &  12.7 & 5.59 & 3.45 & 4.2  \\
2001.38  &  20  &  16  & 2  &  134 &  225 & -19.5 & 5.40 & 2.49 & 3.1  \\
2001.48  &  20  &  16  & 2  &  138 &  238 & -23.6 & 5.31 & 2.57 & 3.2  \\
2001.62  &  20  &  16  & 2  &  137 &  256 & -31.7 & 5.41 & 2.83 & 4.5  \\
2001.70  &  20  &  16  & 2  &  129 &  239 & -28.5 & 4.79 & 2.28 & 3.5  \\
2001.78  &  20  &  16  & 2  &  134 &  248 & -21.3 & 5.69 & 2.63 & 3.6  \\
2001.86  &  20  &  16  & 2  &  135 &  243 & -20.7 & 6.76 & 3.58 & 2.7  \\
2002.07  &  20  &  16  & 2  &  160 &  249 & -20.8 & 7.74 & 3.99 & 2.9  \\
2002.40  &  25  &  16  & 2  &  141 &  212 & -16.0 & 7.75 & 4.27 & 3.0  \\
2002.48  & 220  &  32  & 2  &  139 &  184 &  -3.6 & 7.60 & 3.92 & 0.8  \\
2003.20  & 220  &  32  & 2  &  147 &  186 &   1.1 & 6.19 & 2.60 & 0.9  \\
2003.49  & 220  &  32  & 2  &  140 &  186 &   0.1 & 7.36 & 3.42 & 0.8  \\
2004.64  & 220  &  32  & 2  &  119 &  170 &  -8.3 & 7.02 & 3.22 & 1.1  \\
2004.97  & 220  &  32  & 2  &  129 &  196 & -26.5 & 5.37 & 2.10 & 0.9  \\
2005.25  & 220  &  32  & 2  &  121 &  177 &   3.1 & 4.82 & 1.59 & 1.0  \\
2006.17  &  84  &  64  & 1  &  114 &  190 &   2.7 & 4.27 & 1.91 & 1.5  \\
2006.24  &  84  &  64  & 1  &  113 &  207 &   1.4 & 4.31 & 2.01 & 1.5  \\
2006.30  &  84  &  64  & 1  &  111 &  186 &  -1.0 & 4.34 & 1.91 & 1.6  \\
2006.37  &  84  &  64  & 1  &  110 &  187 &  -3.8 & 4.20 & 1.77 & 1.9  \\
2006.46  &  84  &  64  & 1  &  112 &  188 & -14.8 & 4.21 & 1.83 & 1.9  \\
2006.88  &  42  &  32  & 2  &  130 &  276 & -32.4 & 4.26 & 2.33 & 1.2  \\
2007.16  &  42  &  32  & 2  &  124 &  266 & -31.4 & 6.32 & 3.07 & 1.7  \\
\hline\noalign{\smallskip}
\multicolumn{10}{c}{8.4~GHz images}\\
\hline\noalign{\smallskip}
2002.48  & 84  &  32  & 2  &  1180 &  904 &  -3.6 & 6.73 & 5.65 & 0.9 \\
\noalign{\smallskip}
\hline
\end{tabular}
\end{flushleft}
\label{obslog}
\end{table*}

\addtocounter{table}{1}
% if table 2
\longtab{2}{
\begin{longtable}{lcccc}
\caption{\label{mfits} 86~GHz, 43~GHz and 8.4~GHz circular-Gaussian model-fit parameters. $S$, $r$, $\Theta$ and $FWHM$ are the flux density, the projected distance to the Core, the position angle respect to the Core and the FWHM size, respectively. Upper limits on the $FWHM$ size and $S$ corresponds to fits of unresolved jet regions. For such fits, the upper limit of the sizes were imposed in Difmap to estimate the corresponding flux density upper limits. The latter are close approximations to the actual flux densities of the unresolved regions.}\\
\hline\hline
Comp. &   $S$ &  $r$  &   $\Theta$ &  $FWHM$\\
      & (Jy) & (mas) & ($^{\circ}$) & (mas) \\
\hline
\endfirsthead
\caption{continued.}\\
\hline\hline
Comp. &   $S$ & $r$  &    $\Theta$ &  $FWHM$\\
      & (Jy) & (mas) &  ($^{\circ}$)  & (mas) \\
\hline
\endhead
\hline
\endfoot
\multicolumn{5}{c}{86~GHz  -- 2001.82}\\
\hline\noalign{\smallskip}
Core &  1.937$\pm$ 0.019 &  ... &  ... &  0.053$\pm$ 0.001 \\
W0   &  0.661$\pm$ 0.020 &   0.13$\pm$  0.02 & -109.9$\pm$   9.0 &  0.137$\pm$ 0.007 \\ 
W2   &  $\aplt$0.244 &   0.28$\pm$  0.01 & -140.5$\pm$   2.1 &  $<$0.010 \\ 
W3   &  $\aplt$0.294 &   0.22$\pm$  0.01 & -130.8$\pm$   2.6 &  $<$0.010 \\ 
\hline\noalign{\smallskip}
\multicolumn{5}{c}{86~GHz  -- 2002.30}\\
\hline\noalign{\smallskip}
Core &  1.799$\pm$ 0.018 &   ... &  ... &  0.042$\pm$ 0.000 \\ 
W0   &  0.470$\pm$ 0.047 &   0.18$\pm$  0.02 & -129.2$\pm$   5.3 &  0.034$\pm$ 0.003 \\ 
W2   &  $\aplt$0.111 &   0.31$\pm$  0.01 & -146.2$\pm$   1.8 &  $<$0.010 \\ 
W3   &  $\aplt$0.159 &   0.26$\pm$  0.01 & -139.2$\pm$   2.2 &  $<$0.010 \\ 
\hline\noalign{\smallskip}
\multicolumn{5}{c}{86~GHz  -- 2002.81}\\
\hline\noalign{\smallskip}
Core &  2.332$\pm$ 0.023 &   ...& ... &  0.050$\pm$ 0.001 \\ 
W0   &  0.933$\pm$ 0.028 &   0.15$\pm$  0.02 & -132.9$\pm$   7.8 &  0.106$\pm$ 0.005 \\ 
W2   &  0.278$\pm$ 0.028 &   0.32$\pm$  0.01 & -149.0$\pm$   2.1 &  0.024$\pm$ 0.002 \\ 
W3   &  $\aplt$0.138 &   0.22$\pm$  0.01 & -145.8$\pm$   2.6 &  $<$0.010 \\ 
\hline\noalign{\smallskip}
\multicolumn{5}{c}{86~GHz --  2003.32}\\
\hline\noalign{\smallskip}
Core &  1.376$\pm$ 0.014 &   ... &  ... &  0.059$\pm$ 0.001 \\ 
W2   &  $\aplt$0.224 &   0.30$\pm$  0.01 & -151.6$\pm$   1.9 &  $<$0.010 \\ 
W3   &  0.326$\pm$ 0.033 &   0.22$\pm$  0.02 & -146.1$\pm$   4.8 &  0.037$\pm$ 0.004 \\  
\hline\noalign{\smallskip}
\multicolumn{5}{c}{86~GHz --  2004.29}\\
\hline\noalign{\smallskip}
Core &  3.663$\pm$ 0.037 &   ... &  ... &  0.092$\pm$ 0.001 \\ 
W2   &  0.093$\pm$ 0.009 &   0.37$\pm$  0.01 & -152.4$\pm$   1.7 &  0.021$\pm$ 0.002 \\ 
W3   &  0.785$\pm$ 0.024 &   0.25$\pm$  0.02 & -147.0$\pm$   4.6 &  0.111$\pm$ 0.006 \\ 
\hline\noalign{\smallskip}
\multicolumn{5}{c}{43~GHz --  1997.51}\\
\hline\noalign{\smallskip}
 C &  3.046$\pm$ 0.253 &        ...        &        ...        &  0.085$\pm$ 0.001 \\ 
Q1 &  $\aplt$0.180 &   0.18$\pm$  0.03 &   22.2$\pm$  10.2 &  $<$0.030 \\ 
Q2 &  $\aplt$0.061 &   0.18$\pm$  0.03 &  132.6$\pm$  10.3 &  $<$0.030 \\ 
\hline\noalign{\smallskip}
\multicolumn{5}{c}{43~GHz --  1998.14}\\
\hline\noalign{\smallskip}
 C &  3.051$\pm$ 0.147 &        ...        &        ...        &  0.073$\pm$ 0.001 \\ 
Q1 &  0.573$\pm$ 0.032 &   0.13$\pm$  0.03 &   43.4$\pm$  13.8 &  0.111$\pm$ 0.006 \\ 
Q2 &  $\aplt$0.485 &   0.12$\pm$  0.01 &  118.7$\pm$   4.9 &  $<$0.030 \\ 
\hline\noalign{\smallskip}
\multicolumn{5}{c}{43~GHz --  1998.88}\\
\hline\noalign{\smallskip}
 C &  3.855$\pm$ 0.186 &        ...        &        ...        &  0.106$\pm$ 0.005 \\ 
Q1 &  $\aplt$0.154 &   0.17$\pm$  0.03 &   32.5$\pm$  10.5 &  $<$0.030 \\ 
Q2 &  $\aplt$0.224 &   0.14$\pm$  0.03 &  122.8$\pm$  12.9 &  $<$0.030 \\ 
\hline\noalign{\smallskip}
\multicolumn{5}{c}{43~GHz --  1999.10}\\
\hline\noalign{\smallskip}
 C &  3.534$\pm$ 0.120 &        ...        &        ...        &  0.066$\pm$ 0.001 \\ 
Q1 &  0.602$\pm$ 0.027 &   0.13$\pm$  0.03 &   50.5$\pm$  14.0 &  0.101$\pm$ 0.005 \\ 
Q2 &  $\aplt$0.653 &   0.13$\pm$  0.01 &  121.1$\pm$   4.4 &  $<$0.030 \\ 
\hline\noalign{\smallskip}
\multicolumn{5}{c}{43~GHz --  2000.54}\\
\hline\noalign{\smallskip}
 C &  3.550$\pm$ 0.243 &        ...        &        ...        &  0.084$\pm$ 0.001 \\ 
Q1 &  0.685$\pm$ 0.051 &   0.13$\pm$  0.03 &   49.3$\pm$  13.8 &  0.136$\pm$ 0.007 \\ 
Q2 &  1.480$\pm$ 0.101 &   0.15$\pm$  0.01 &  128.5$\pm$   3.7 &  0.057$\pm$ 0.001 \\ 
\hline\noalign{\smallskip}
\multicolumn{5}{c}{43~GHz --  2000.72}\\
\hline\noalign{\smallskip}
 C &  3.645$\pm$ 0.154 &        ...        &        ...        &  0.087$\pm$ 0.001 \\ 
Q1 &  0.619$\pm$ 0.031 &   0.15$\pm$  0.03 &   50.2$\pm$  12.5 &  0.133$\pm$ 0.007 \\ 
Q2 &  1.794$\pm$ 0.076 &   0.16$\pm$  0.01 &  130.5$\pm$   3.5 &  0.078$\pm$ 0.001 \\ 
\hline\noalign{\smallskip}
\multicolumn{5}{c}{43~GHz --  2000.80}\\
\hline\noalign{\smallskip}
 C &  3.562$\pm$ 0.187 &        ...        &        ...        &  0.085$\pm$ 0.001 \\ 
Q1 &  0.838$\pm$ 0.050 &   0.13$\pm$  0.03 &   53.7$\pm$  14.1 &  0.160$\pm$ 0.008 \\ 
Q2 &  1.744$\pm$ 0.091 &   0.17$\pm$  0.01 &  130.4$\pm$   3.3 &  0.080$\pm$ 0.001 \\ 
\hline\noalign{\smallskip}
\multicolumn{5}{c}{43~GHz --  2000.88}\\
\hline\noalign{\smallskip}
 C &  3.801$\pm$ 0.177 &        ...        &        ...        &  0.088$\pm$ 0.001 \\ 
Q1 &  $\aplt$0.430 &   0.17$\pm$  0.03 &   60.1$\pm$  11.1 &  $<$0.030 \\ 
Q2 &  1.944$\pm$ 0.106 &   0.18$\pm$  0.03 &  133.6$\pm$  10.1 &  0.101$\pm$ 0.005 \\ 
\hline\noalign{\smallskip}
\multicolumn{5}{c}{43~GHz --  2000.97}\\
\hline\noalign{\smallskip}
 C &  3.885$\pm$ 0.145 &        ...        &        ...        &  0.092$\pm$ 0.001 \\ 
Q1 &  $\aplt$0.576 &   0.16$\pm$  0.03 &   70.2$\pm$  11.6 &  $<$0.030 \\ 
Q2 &  1.461$\pm$ 0.055 &   0.20$\pm$  0.01 &  138.4$\pm$   2.9 &  0.067$\pm$ 0.001 \\ 
\hline\noalign{\smallskip}
\multicolumn{5}{c}{43~GHz --  2001.05}\\
\hline\noalign{\smallskip}
 C &  3.618$\pm$ 0.111 &        ...        &        ...        &  0.085$\pm$ 0.001 \\ 
Q1 &  0.541$\pm$ 0.023 &   0.17$\pm$  0.03 &   56.4$\pm$  11.1 &  0.058$\pm$ 0.003 \\ 
Q2 &  1.581$\pm$ 0.048 &   0.18$\pm$  0.01 &  131.6$\pm$   3.1 &  0.055$\pm$ 0.001 \\ 
\hline\noalign{\smallskip}
\multicolumn{5}{c}{43~GHz --  2001.13}\\
\hline\noalign{\smallskip}
 C &  2.889$\pm$ 0.085 &        ...        &        ...        &  0.087$\pm$ 0.001 \\ 
Q1 &  1.084$\pm$ 0.044 &   0.12$\pm$  0.03 &   71.4$\pm$  14.6 &  0.148$\pm$ 0.007 \\ 
Q2 &  1.766$\pm$ 0.052 &   0.20$\pm$  0.01 &  135.4$\pm$   2.9 &  0.098$\pm$ 0.001 \\ 
\hline\noalign{\smallskip}
\multicolumn{5}{c}{43~GHz --  2001.20}\\
\hline\noalign{\smallskip}
 C &  3.733$\pm$ 0.149 &        ...        &        ...        &  0.120$\pm$ 0.006 \\ 
Q2 &  2.001$\pm$ 0.080 &   0.19$\pm$  0.03 &  135.9$\pm$   9.5 &  0.114$\pm$ 0.006 \\ 
\hline\noalign{\smallskip}
\multicolumn{5}{c}{43~GHz --  2001.29}\\
\hline\noalign{\smallskip}
 C &  3.097$\pm$ 0.104 &        ...        &        ...        &  0.089$\pm$ 0.001 \\ 
Q1 &  0.662$\pm$ 0.029 &   0.15$\pm$  0.03 &   73.7$\pm$  12.4 &  0.142$\pm$ 0.007 \\ 
Q2 &  1.815$\pm$ 0.061 &   0.20$\pm$  0.01 &  135.8$\pm$   2.9 &  0.084$\pm$ 0.001 \\ 
\hline\noalign{\smallskip}
\multicolumn{5}{c}{43~GHz --  2001.38}\\
\hline\noalign{\smallskip}
 C &  2.781$\pm$ 0.107 &        ...        &        ...        &  0.089$\pm$ 0.001 \\ 
Q1 &  0.439$\pm$ 0.021 &   0.17$\pm$  0.03 &   66.3$\pm$  10.7 &  0.107$\pm$ 0.005 \\ 
Q2 &  2.159$\pm$ 0.083 &   0.19$\pm$  0.01 &  136.8$\pm$   3.0 &  0.089$\pm$ 0.001 \\ 
\hline\noalign{\smallskip}
\multicolumn{5}{c}{43~GHz --  2001.48}\\
\hline\noalign{\smallskip}
 C &  2.692$\pm$ 0.104 &        ...        &        ...        &  0.091$\pm$ 0.001 \\ 
Q1 &  0.344$\pm$ 0.016 &   0.18$\pm$  0.03 &   68.3$\pm$  10.2 &  0.065$\pm$ 0.003 \\ 
Q2 &  2.219$\pm$ 0.086 &   0.19$\pm$  0.01 &  138.9$\pm$   3.0 &  0.097$\pm$ 0.001 \\ 
\hline\noalign{\smallskip}
\multicolumn{5}{c}{43~GHz --  2001.62}\\
\hline\noalign{\smallskip}
 C &  2.623$\pm$ 0.123 &        ...        &        ...        &  0.082$\pm$ 0.001 \\ 
Q1 &  0.456$\pm$ 0.025 &   0.17$\pm$  0.03 &   67.6$\pm$  10.6 &  0.108$\pm$ 0.005 \\ 
Q2 &  2.260$\pm$ 0.106 &   0.20$\pm$  0.01 &  140.1$\pm$   2.9 &  0.096$\pm$ 0.001 \\ 
\hline\noalign{\smallskip}
\multicolumn{5}{c}{43~GHz --  2001.70}\\
\hline\noalign{\smallskip}
 C &  2.440$\pm$ 0.159 &        ...        &        ...        &  0.089$\pm$ 0.001 \\ 
Q1 &  $\aplt$0.250 &   0.19$\pm$  0.03 &   67.6$\pm$   9.6 &  $<$0.030 \\ 
Q2 &  2.068$\pm$ 0.135 &   0.20$\pm$  0.01 &  139.6$\pm$   2.8 &  0.093$\pm$ 0.001 \\ 
\hline\noalign{\smallskip}
\multicolumn{5}{c}{43~GHz --  2001.78}\\
\hline\noalign{\smallskip}
 C &  3.036$\pm$ 0.173 &        ...        &        ...        &  0.095$\pm$ 0.001 \\ 
Q1 &  0.437$\pm$ 0.028 &   0.20$\pm$  0.03 &   71.5$\pm$   9.3 &  0.106$\pm$ 0.005 \\ 
Q2 &  2.184$\pm$ 0.125 &   0.21$\pm$  0.01 &  137.7$\pm$   2.7 &  0.090$\pm$ 0.001 \\ 
\hline\noalign{\smallskip}
\multicolumn{5}{c}{43~GHz --  2001.86}\\
\hline\noalign{\smallskip}
 C &  3.600$\pm$ 0.162 &        ...        &        ...        &  0.054$\pm$ 0.001 \\ 
Q1 &  0.594$\pm$ 0.032 &   0.20$\pm$  0.03 &   84.4$\pm$   9.2 &  0.186$\pm$ 0.009 \\ 
Q2 &  1.625$\pm$ 0.073 &   0.25$\pm$  0.01 &  138.4$\pm$   2.3 &  0.063$\pm$ 0.001 \\ 
Q3 &  0.944$\pm$ 0.042 &   0.15$\pm$  0.01 &  117.6$\pm$   3.9 &  0.031$\pm$ 0.001 \\ 
\hline\noalign{\smallskip}
\multicolumn{5}{c}{43~GHz --  2002.07}\\
\hline\noalign{\smallskip}
 C &  4.334$\pm$ 0.145 &       ...         &        ...        &  0.086$\pm$ 0.001 \\ 
Q1 &  0.790$\pm$ 0.035 &   0.16$\pm$  0.03 &   83.9$\pm$  11.4 &  0.201$\pm$ 0.010 \\ 
Q2 &  1.508$\pm$ 0.050 &   0.26$\pm$  0.01 &  143.0$\pm$   2.2 &  0.100$\pm$ 0.001 \\ 
Q3 &  $\aplt$1.069 &   0.19$\pm$  0.01 &  125.6$\pm$   3.0 &  $<$0.030 \\ 
\hline\noalign{\smallskip}
\multicolumn{5}{c}{43~GHz --  2002.40}\\
\hline\noalign{\smallskip}
 C &  4.204$\pm$ 0.177 &        ...        &        ...        &  0.046$\pm$ 0.001 \\ 
Q1 &  0.884$\pm$ 0.045 &   0.18$\pm$  0.03 &   92.9$\pm$   9.9 &  0.298$\pm$ 0.015 \\ 
Q2 &  0.921$\pm$ 0.039 &   0.28$\pm$  0.01 &  144.4$\pm$   2.0 &  0.032$\pm$ 0.001 \\ 
Q3 &  1.695$\pm$ 0.086 &   0.17$\pm$  0.03 &  127.8$\pm$  11.0 &  0.102$\pm$ 0.005 \\ 
\hline\noalign{\smallskip}
\multicolumn{5}{c}{43~GHz --  2002.48}\\
\hline\noalign{\smallskip}
 C &  4.409$\pm$ 0.171 &        ...        &        ...        &  0.058$\pm$ 0.001 \\ 
Q1 &  0.576$\pm$ 0.028 &   0.21$\pm$  0.03 &   86.6$\pm$   8.7 &  0.216$\pm$ 0.011 \\ 
Q2 &  1.045$\pm$ 0.040 &   0.28$\pm$  0.01 &  144.0$\pm$   2.1 &  0.060$\pm$ 0.001 \\ 
Q3 &  1.540$\pm$ 0.074 &   0.17$\pm$  0.03 &  129.8$\pm$  11.1 &  0.103$\pm$ 0.005 \\ 
\hline\noalign{\smallskip}
\multicolumn{5}{c}{43~GHz --  2003.20}\\
\hline\noalign{\smallskip}
 C &  3.356$\pm$ 0.145 &        ...        &        ...        &  0.085$\pm$ 0.001 \\ 
Q1 &  0.582$\pm$ 0.030 &   0.23$\pm$  0.03 &  101.0$\pm$   8.1 &  0.180$\pm$ 0.009 \\ 
Q2 &  0.788$\pm$ 0.034 &   0.32$\pm$  0.01 &  150.8$\pm$   1.8 &  0.050$\pm$ 0.001 \\ 
Q3 &  1.440$\pm$ 0.074 &   0.19$\pm$  0.03 &  146.7$\pm$   9.9 &  0.107$\pm$ 0.005 \\ 
\hline\noalign{\smallskip}
\multicolumn{5}{c}{43~GHz --  2003.49}\\
\hline\noalign{\smallskip}
 C &  4.647$\pm$ 0.259 &        ...        &        ...        &  0.088$\pm$ 0.001 \\ 
Q1 &  0.554$\pm$ 0.035 &   0.24$\pm$  0.03 &   95.6$\pm$   7.7 &  0.204$\pm$ 0.010 \\ 
Q2 &  $\aplt$0.569 &   0.31$\pm$  0.01 &  151.5$\pm$   1.8 &  $<$0.030 \\ 
Q3 &  1.596$\pm$ 0.089 &   0.19$\pm$  0.01 &  148.5$\pm$   3.0 &  0.097$\pm$ 0.001 \\ 
\hline\noalign{\smallskip}
\multicolumn{5}{c}{43~GHz --  2004.64}\\
\hline\noalign{\smallskip}
 C &  4.937$\pm$ 0.160 &        ...        &        ...        &  0.107$\pm$ 0.005 \\ 
Q1 &  0.558$\pm$ 0.018 &   0.30$\pm$  0.03 &  108.0$\pm$   6.1 &  0.229$\pm$ 0.011 \\ 
Q2 &  0.438$\pm$ 0.007 &   0.38$\pm$  0.01 &  155.6$\pm$   1.5 &  0.035$\pm$ 0.000 \\ 
Q3 &  1.063$\pm$ 0.017 &   0.23$\pm$  0.01 &  151.8$\pm$   2.5 &  0.098$\pm$ 0.001 \\ 
\hline\noalign{\smallskip}
\multicolumn{5}{c}{43~GHz --  2004.97}\\
\hline\noalign{\smallskip}
 C &  3.229$\pm$ 0.116 &        ...        &        ...        &  0.128$\pm$ 0.006 \\ 
Q1 &  0.475$\pm$ 0.017 &   0.31$\pm$  0.03 &  110.6$\pm$   6.0 &  0.204$\pm$ 0.010 \\ 
Q2 &  0.507$\pm$ 0.018 &   0.40$\pm$  0.03 &  158.1$\pm$   4.6 &  0.069$\pm$ 0.003 \\ 
Q3 &  1.160$\pm$ 0.042 &   0.21$\pm$  0.03 &  157.6$\pm$   8.6 &  0.117$\pm$ 0.006 \\ 
\hline\noalign{\smallskip}
\multicolumn{5}{c}{43~GHz --  2005.25}\\
\hline\noalign{\smallskip}
 C &  2.512$\pm$ 0.093 &        ...        &        ...        &  0.119$\pm$ 0.006 \\ 
Q1 &  0.435$\pm$ 0.016 &   0.33$\pm$  0.03 &  111.7$\pm$   5.6 &  0.226$\pm$ 0.011 \\ 
Q2 &  $\aplt$0.875 &   0.40$\pm$  0.01 &  160.3$\pm$   1.4 &  $<$0.030 \\ 
Q3 &  1.009$\pm$ 0.037 &   0.22$\pm$  0.03 &  162.3$\pm$   8.6 &  0.153$\pm$ 0.008 \\ 
\hline\noalign{\smallskip}
\multicolumn{5}{c}{43~GHz --  2006.17}\\
\hline\noalign{\smallskip}
 C &  1.280$\pm$ 0.049 &        ...        &        ...        &  0.148$\pm$ 0.007 \\ 
Q1 &  0.261$\pm$ 0.010 &   0.36$\pm$  0.03 &  115.6$\pm$   5.1 &  0.217$\pm$ 0.011 \\ 
Q2 &  1.788$\pm$ 0.045 &   0.36$\pm$  0.01 &  160.2$\pm$   1.6 &  0.039$\pm$ 0.000 \\ 
Q3 &  0.940$\pm$ 0.024 &   0.25$\pm$  0.01 &  164.8$\pm$   2.3 &  0.087$\pm$ 0.001 \\ 
\hline\noalign{\smallskip}
\multicolumn{5}{c}{43~GHz --  2006.24}\\
\hline\noalign{\smallskip}
 C &  1.257$\pm$ 0.049 &        ...        &        ...        &  0.142$\pm$ 0.007 \\ 
Q1 &  0.349$\pm$ 0.014 &   0.34$\pm$  0.03 &  113.7$\pm$   5.3 &  0.233$\pm$ 0.012 \\ 
Q2 &  $\aplt$1.484 &   0.36$\pm$  0.01 &  160.4$\pm$   1.6 &  $<$0.030 \\ 
Q3 &  1.218$\pm$ 0.033 &   0.26$\pm$  0.01 &  163.5$\pm$   2.2 &  0.082$\pm$ 0.001 \\ 
\hline\noalign{\smallskip}
\multicolumn{5}{c}{43~GHz --  2006.30}\\
\hline\noalign{\smallskip}
 C &  1.207$\pm$ 0.049 &        ...        &        ...        &  0.153$\pm$ 0.008 \\ 
Q1 &  0.362$\pm$ 0.015 &   0.34$\pm$  0.03 &  112.1$\pm$   5.3 &  0.248$\pm$ 0.012 \\ 
Q2 &  1.470$\pm$ 0.043 &   0.35$\pm$  0.01 &  160.2$\pm$   1.6 &  0.051$\pm$ 0.001 \\ 
Q3 &  1.295$\pm$ 0.038 &   0.25$\pm$  0.01 &  163.5$\pm$   2.2 &  0.077$\pm$ 0.001 \\ 
\hline\noalign{\smallskip}
\multicolumn{5}{c}{43~GHz --  2006.37}\\
\hline\noalign{\smallskip}
 C &  1.133$\pm$ 0.047 &        ...        &        ...        &  0.155$\pm$ 0.008 \\ 
Q1 &  0.390$\pm$ 0.016 &   0.33$\pm$  0.03 &  115.8$\pm$   5.6 &  0.262$\pm$ 0.013 \\ 
Q2 &  1.236$\pm$ 0.037 &   0.35$\pm$  0.01 &  160.6$\pm$   1.6 &  0.056$\pm$ 0.001 \\ 
Q3 &  1.430$\pm$ 0.043 &   0.25$\pm$  0.01 &  163.4$\pm$   2.3 &  0.077$\pm$ 0.001 \\ 
\hline\noalign{\smallskip}
\multicolumn{5}{c}{43~GHz --  2006.46}\\
\hline\noalign{\smallskip}
 C &  1.079$\pm$ 0.045 &        ...        &        ...        &  0.149$\pm$ 0.007 \\ 
Q1 &  0.338$\pm$ 0.014 &   0.35$\pm$  0.03 &  114.6$\pm$   5.3 &  0.258$\pm$ 0.013 \\ 
Q2 &  1.038$\pm$ 0.032 &   0.36$\pm$  0.01 &  161.0$\pm$   1.6 &  0.049$\pm$ 0.000 \\ 
Q3 &  1.742$\pm$ 0.054 &   0.25$\pm$  0.01 &  162.8$\pm$   2.3 &  0.079$\pm$ 0.001 \\ 
\hline\noalign{\smallskip}
\multicolumn{5}{c}{43~GHz --  2006.88}\\
\hline\noalign{\smallskip}
 C &  0.760$\pm$ 0.027 &        ...        &        ...        &  0.150$\pm$ 0.008 \\ 
Q1 &  0.253$\pm$ 0.009 &   0.38$\pm$  0.03 &  117.5$\pm$   4.8 &  0.274$\pm$ 0.014 \\ 
Q2 &  0.769$\pm$ 0.017 &   0.42$\pm$  0.01 &  163.2$\pm$   1.4 &  0.049$\pm$ 0.000 \\ 
Q3 &  2.465$\pm$ 0.055 &   0.24$\pm$  0.01 &  163.4$\pm$   2.4 &  0.076$\pm$ 0.001 \\ 
\hline\noalign{\smallskip}
\multicolumn{5}{c}{43~GHz --  2007.16}\\
\hline\noalign{\smallskip}
 C &  0.903$\pm$ 0.052 &        ...        &        ...        &  0.170$\pm$ 0.009 \\ 
Q1 &  0.348$\pm$ 0.020 &   0.38$\pm$  0.03 &  119.7$\pm$   4.8 &  0.258$\pm$ 0.013 \\ 
Q2 &  1.143$\pm$ 0.057 &   0.45$\pm$  0.01 &  163.8$\pm$   1.3 &  0.085$\pm$ 0.001 \\ 
Q3 &  3.895$\pm$ 0.222 &   0.23$\pm$  0.03 &  164.0$\pm$   7.9 &  0.101$\pm$ 0.005 \\ 
\hline\noalign{\smallskip}
\multicolumn{5}{c}{8.4~GHz  -- 2002.48}\\
\hline\noalign{\smallskip}
Core &  6.352$\pm$ 0.635 &   ... &   ... &  0.378$\pm$ 0.038 \\ 
X4   &  $\aplt$0.088 &   0.62$\pm$  0.24 &  -98.8$\pm$  22.3 &  $<$0.180 \\ 
X5   &  0.073$\pm$ 0.007 &   1.73$\pm$  0.42 &  -52.5$\pm$  13.9 &  0.836$\pm$ 0.084 \\ 
X6   &  0.116$\pm$ 0.012 &   3.59$\pm$  0.44 &  -43.0$\pm$   7.1 &  0.883$\pm$ 0.088 \\ 
X7   &  0.103$\pm$ 0.010 &  24.71$\pm$  1.39 &  -31.8$\pm$   3.2 &  2.785$\pm$ 0.278 \\ 
\end{longtable}
}% End \longtab

\begin{table*}
\centering
\caption[]{ Proper motion results.
For each model component, we give the number of data points for the fit of the trajectory ($N$),
the mean position angle ($<\Theta>$), the mean angular speed ($<\dot{\Theta}>$), the mean proper motion ($<\mu>$, and its corresponding $\beta^{\rm{obs}}_{\rm{app}}$$<\mu>$), as well as its radial ($<\mu_{\rm{rad}}>$) and normal component ($<\mu_{\rm{norm}}>$, and its corresponding  $\beta^{\rm{obs}}_{\rm{app\_norm}}$).}
\begin{flushleft}
\begin{tabular} {lcccccccc}
\hline\noalign{\smallskip}
Comp. &  $N$  & $<\Theta>$    & $<\dot{\Theta}>$ & $<\mu>$  &    $\beta^{\rm{obs}}_{\rm{app}}$    & $<\mu_{\rm{rad}}>$ & $<\mu_{\rm{norm}}>$  &  $\beta^{\rm{obs}}_{\rm{app\_norm}}$    \\
     &     & $^{\circ}$    & $^{\circ}$/yr &   mas/yr &                 $c$                 &    mas/yr       &     mas/yr     &   $c$        \\
\hline\noalign{\smallskip}
Q1   &   33  &  139.5$\pm$7.1  &  10.7$\pm$0.7 & 0.047$\pm$0.002 &  3.26$\pm$0.14  & 0.026$\pm$0.002  &  0.039$\pm$0.003 &   2.71$\pm$0.21 \\
Q2   &   34  &  175.4$\pm$1.3  &   5.8$\pm$0.3 & 0.042$\pm$0.001 &  2.85$\pm$0.07  & 0.036$\pm$0.001  &  0.020$\pm$0.003 &   1.39$\pm$0.21 \\
Q3   &   16  &  153.8$\pm$2.7  &   8.0$\pm$0.5 & 0.034$\pm$0.002 &  2.29$\pm$0.14  & 0.018$\pm$0.002  &  0.029$\pm$0.002 &   2.01$\pm$0.14 \\ 
\noalign{\smallskip}
\hline
\end{tabular}
\end{flushleft}
\label{kintab}
\end{table*}

\end{document}